\documentclass[iop, tighten]{emulateapj}

\usepackage{times}
\usepackage{epsfig}
\usepackage{colordvi}
\usepackage{graphicx}
\usepackage{amssymb}

\def\apj{ApJ}
\def\apjl{ApJ}
\def\apjs{ApJS}
\def\aap{A\&A}
\def\mnras{MNRAS}
\def\pasj{PASJ}
\def\aaps{A\&AS}
\def\aj{AJ}
\def\physrep{Phys. Rep.}
 
\newcommand{\be}{\begin{equation}}
\newcommand{\ee}{\end{equation}}
\newcommand{\bea}{\begin{eqnarray}}
\newcommand{\eea}{\end{eqnarray}}
\newcommand{\rxte}{{\it RXTE}}
\newcommand{\source}{\mbox{4U\,1724--307}}
\newcommand{\msun}{{M}_{\odot}}

\begin{document}

   \title{Neutron star stiff equation of state derived from cooling phases of the X-ray burster 4U~1724$-$307}

\shorttitle{Cooling phases of the X-ray burster 4U~1724$-$307}
\shortauthors{Suleimanov et al.}

\author{Valery Suleimanov, \altaffilmark{1,2}  
Juri~Poutanen,\altaffilmark{3} Mikhail  Revnivtsev,\altaffilmark{4} and    Klaus Werner\altaffilmark{1} }

\affil{$^1$Institute for Astronomy and Astrophysics, Kepler Center for Astro and
Particle Physics, Eberhard Karls University, Sand 1,
 72076 T\"ubingen, Germany; [suleimanov,werner]@astro.uni-tuebingen.de  \\
$^2$Kazan Federal University, Kremlevskaja str., 18, Kazan 420008, Russia \\
$^3$ Astronomy Division, Department of Physics, PO Box 3000, FI-90014 University of Oulu, Finland; juri.poutanen@oulu.fi \\
$^4$ Space Research Institute (IKI), Russian Academy of Sciences, Profsoyuznaya 84/32, 117997 Moscow, Russia}

\submitted{Accepted to ApJ on 2011 September 5}
%\date{Accepted 2007 November 11. Received 2007 October 29; in original form 2007 September 19}
%\pagerange{\pageref{firstpage}--\pageref{lastpage}} \pubyear{2011}
%\date{}
%\pubyear{2011}
%\maketitle
%\label{firstpage}

 \begin{abstract} 
 Thermal emission during X-ray bursts is a powerful tool to determine neutron
 star masses and radii, if the Eddington flux and 
the apparent radius in the cooling tail  can be measured accurately, and distances to the sources are known.  
We propose here an improved method of determining the basic stellar parameters using the data from the cooling phase of
photospheric radius expansion bursts covering a large range of luminosities.  
 Because at that phase the blackbody apparent radius depends only on the spectral hardening factor (color-correction),
we suggest to fit the theoretical dependences  of the
color-correction versus flux in Eddington units to the observed 
variations of the inverse square root of the apparent blackbody radius with the flux.
For that we use a large set of atmosphere models for burst luminosities varying by three orders of magnitude and  
for various chemical compositions and surface gravities.  
We show that   spectral variations observed during a long photospheric radius expansion burst from \source\  
are entirely consistent with the theoretical expectations for the passively cooling neutron star atmospheres. 
Our method allows us to determine both the Eddington flux  
(which is found to be smaller than the touchdown flux by 15\%) 
and the ratio of the stellar apparent radius to the distance much more reliably. 
We then find a lower limit on the neutron star radius of 14 km for masses
below 2.2$\msun$, independently of the chemical 
composition. These results suggest that the matter inside neutron stars is
characterized by a stiff equation of state. 
We also find evidences in favour of hydrogen rich accreting matter and obtain an upper limit to the distance of 7 kpc. 
We  finally show that  the apparent blackbody emitting area  in the cooling tails of the short bursts from 
\source\ is two times smaller  than that for the long burst and their evolution  does not follow the theory.
This makes their usage  for determination of the neutron star parameters questionable and 
casts serious doubts on the results of previous works that used  for the analysis similar bursts from other sources.
\end{abstract}


%________________________________________________________________
\section{Introduction} 

Studies of the thermal emission from neutron stars (NSs) have been used
extensively to determine their masses and radii 
\citep[e.g.][]{Damen90,vP90,LvPT93,R02, HR06,WB07,OGP09,GO10}, which can
provide constraints on 
the properties of the matter at supranuclear densities \citep{HPY07,LP07}. 
Thermonuclear X-ray bursts at NS surfaces and, particularly, the photospheric radius expansion (PRE) bursts 
exceeding the Eddington limit at high-flux phases \citep{Kuulkers03} are excellent laboratories for such detailed studies.
The observed Eddington flux gives a constraint on the NS mass-radius
relation. This method, however, 
suffers from the uncertainty in determination of the exact moment when the
luminosity 
reaches the Eddington limit at the NS surface, which is often assumed to
coincide with 
the moment of ``touchdown'', when  the measured color temperature is highest and the apparent radius is smallest \citep{Damen90}. 
This interpretation is uncertain, as the touchdown flux typically  coincides with the  maximum flux \citep{GOP08}, 
while the latter is expected to be larger  than the surface Eddington flux by the redshift factor $1+z$ \citep{LvPT93}.

The second constraint can be obtained from the apparent radius of the NS at
late stages of the burst. 
This method also has systematic uncertainties related to the color-correction
factor $f_{\rm c}=T_{\rm c}/T_{\rm eff}$ 
(the ratio of the color temperature to the effective temperature of the star),  which is a function of the burst luminosity. 

To minimize the theoretical  uncertainties in modeling the burst atmospheres
at nearly Eddington luminosities, 
we proposed \citep{SPW11} 
to use the whole cooling track and check that the evolution of the blackbody normalization with flux is consistent with the 
theoretically predicted evolution for a passively cooling neutron star.  

The unique determination of   mass and radius requires one more constraint. 
Knowing the distance to a burster breaks the degeneracy, and therefore bursters located in globular clusters would be of  interest. 
For the analysis in the present paper we choose \source\ which resides in the globular cluster Terzan 2.
We analyze three PRE bursts from that source and show that they follow different cooling tracks. 
We also speculate on the reasons for such a discrepancy. 
 We demonstrate that 
the apparent blackbody radius for the long PRE burst  evolves according to the predictions 
for the passively cooling neutron star with a constant apparent surface 
and use these data to determine the NS parameters. 

\newpage 

\section{Data} 

\subsection{X-ray bursts from \source}
\label{sec:data}

For our analysis we have used the data from the PCA spectrometer of {\it Rossi X-ray Timing Explorer (\rxte)},
because it provides us the maximum possible number of photons within the small time intervals, during which 
the spectrum of the X-ray burst does not vary much. 
\rxte/PCA data were analyzed with the help of {\sc heasoft} package (version 6.6.1). 
Response matrices were generated using task {\sc pcarsp} (v.10.1) of this package. 
 The background of PCA detectors was estimated with the help of \verb|CM_bright_VLE| model.
 In order to account for the uncertainties reflecting the quality of the \rxte/PCA response matrix, a 
1\% systematic error  \citep{JMR06}
was added in quadrature to the statistical error in each PCA energy channel.
The spectral  fitting was performed using the {\sc xspec 11} package \citep{Arn96}.

\begin{figure}
\plotone{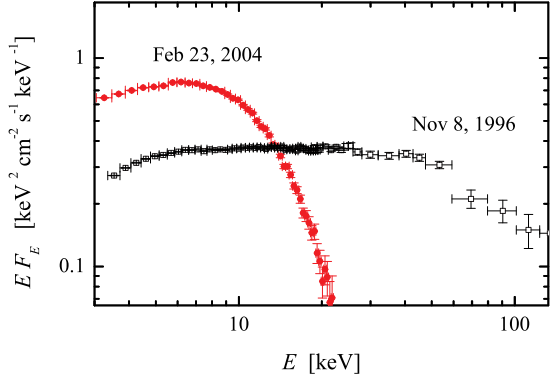}
%\centering 
%\includegraphics[width=0.95\columnwidth]{fig1.eps}
\caption{Spectra of the persistent emission before the long burst (open squares) and one of the short bursts (filled circles).}	
\label{fig:sp_pers}
\end{figure}

\rxte\ observed three PRE bursts from \source\ \citep{GMH08}. 
A long ($>$150 s) PRE burst was recorded on November 8, 1996 \citep{MGL00}.
Two short PRE bursts were observed on Feb 23 and May 22, 2004. The quiescent
emission around these three bursts was significantly different. 
The long burst happened when the source was in the so called hard/low state
with a luminosity of 
a few percent of the Eddington luminosity and the X-ray emission was formed in an optically thin medium. 
The short bursts were observed during the high/soft state, with the X-ray emission  characterized by 
the optically thick accretion disk and the boundary/spreading layer from the neutron star surface.
The quiescent spectra of \source\ are presented in Fig.\,\ref{fig:sp_pers}.
All spectra presented here and below were fitted assuming an absorption column density of $N_{\rm H} = 10^{22}$~cm$^{-2}$
corresponding to the best-fit value for the persistent spectrum. 
But fitting using $N_{\rm H}$ as a free parameter was also performed.

\begin{figure}
\plotone{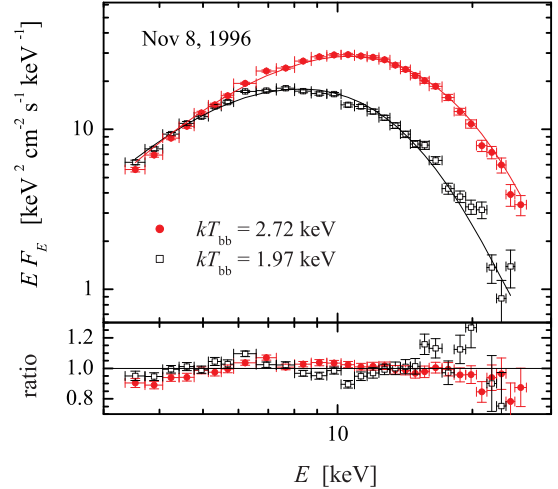}
%\centering 
%\includegraphics[width=0.95\columnwidth]{fig2.eps}
\caption{Evolution of the long burst spectra   at early decline burst phases with the 
corresponding best-fit blackbody models (solid curves). The corresponding time points are shown 
by arrows in the middle panel of Fig.\,\ref{fig:obs}. }	
\label{fig:sp_burst8}
\end{figure}

\begin{figure}
\plotone{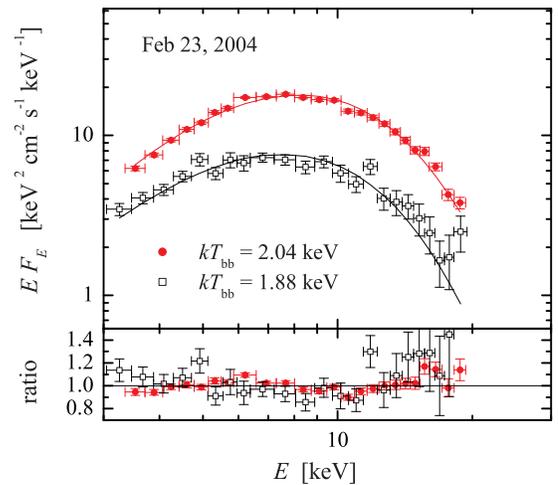}
%\centering 
%\includegraphics[width=0.95\columnwidth]{fig3.eps}
\caption{Same as Fig. \ref{fig:sp_burst8}, but for a short  burst.  }	
\label{fig:sp_burst23}
\end{figure}

X-ray bursts often have thermally looking spectra which can be approximated by a blackbody \citep{GMH08}. 
The fits of the  burst spectra with an absorbed blackbody model give, however, rather high reduced $\chi^2$/d.o.f$\sim$2--3.
This is not very surprising because the photon statistics is high with the errors being dominated by the systematics. 
Theoretical models of the NS atmosphere during X-ray bursts do not predict
exactly blackbody spectral shapes, 
but the residuals  in the 3--20 keV  \rxte\ energy band are expected at the level of 4--5\% (see Fig. 8 in \citealt{SPW11}). 
The residuals between the blackbody model and the data of X-ray bursts of
\source\ have similar amplitudes 
(see Figs.\,\ref{fig:sp_burst8} and \ref{fig:sp_burst23}). Adopting the systematic uncertainty of the model at the level 
 of 3--5\% allows us to obtain acceptable $\chi^2$ values for all spectra at the cooling tails of the analyzed X-ray bursts,
while simple blackbody analytical model allows us to correctly represent the general shape of the spectrum in 3--20 keV energy band. 
Therefore in our subsequent analysis we use simple blackbody models for approximating the spectral shapes and 
compare the obtained parameters with those, 
obtained by fitting the spectra produced by the full theoretical NS atmosphere model in the same energy band.

The best-fit parameters of the employed blackbody model are the color temperature $T_{\rm bb}$ and the normalization constant 
$K\equiv(R_{\rm bb} [{\rm km}]/D_{10})^2$ (the blackbody radius is measured in
km and the distance  $D$ to the source in units of 10 kpc).
These can be combined to estimate the observed bolometric flux $F$ .

\begin{figure}
\plotone{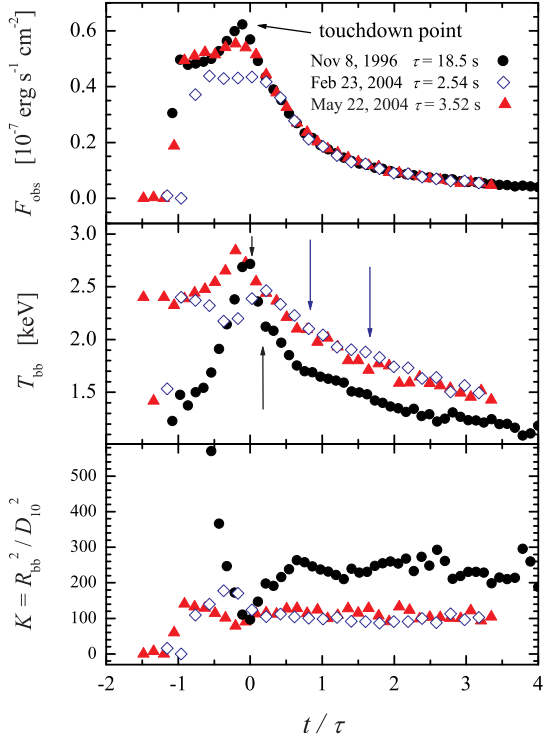}
\centering 
\caption{Evolution of the observed blackbody fluxes, color temperatures and normalizations $K=R^2_{\rm bb}/D_{10}^2$ for three
bursts from \source\ in November 8, 1996 (black circles), February 23, 2004 (blue diamonds) and May 22, 2004 (red triangles).
The time variable is normalized to the characteristic decay time $\tau$ which is equal to 18.5, 2.54 and 3.52 s 
for these three bursts, respectively.  
 For the long burst, the zero time corresponds to the touchdown point (marked by an arrow in the upper and middle panels),
while for the short burst the light curves were shifted so that the cooling tails coincide.  
The arrows show the times when the spectra shown in Figs. \ref{fig:sp_burst8} and 
\ref{fig:sp_burst23} were collected. 
}	
\label{fig:obs}
\end{figure}

The evolution of the fitted parameters during the bursts are shown in Fig.\,\ref{fig:obs}. The time is normalized to the individual 
flux decay timescales $\tau$ and shifted to allow an easy comparison between
the bursts.  The time point, when the 
flux and the color temperature $T_{\rm bb}$ reach their maximum values and the
normalization $K$ has a minimum, 
is usually named the ``touchdown'' point 
 (shown by the arrow in the upper  and middle  panels of Fig.\,\ref{fig:obs} for the long burst). 
The maximal fluxes of the short bursts are appreciably smaller than that for the long burst, and their 
normalizations do not show a significant rise at the early stages of the bursts indicating that the NS 
photosphere has not substantially expanded.
  
\begin{figure}
\plotone{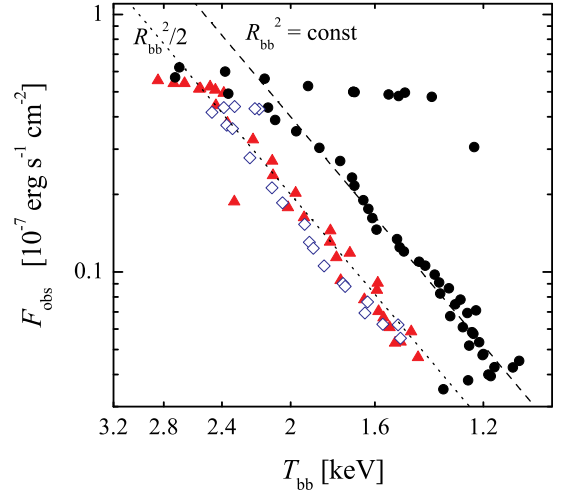}
%\centering 
%\includegraphics[width=0.95\columnwidth]{fig5.eps}
\caption{Evolution of the bursts in the flux -- temperature plane for the three bursts from Fig.\,\ref{fig:obs}. 
The curves of constant $R_{\rm bb}^2$ describing the decay phase are also shown. The blackbody apparent area for the short bursts 
(dotted curve) is a factor of two smaller than the corresponding area describing the long burst (dashed curve). }	
\label{fig:HR}
\end{figure}

The most serious difference between the short and the long bursts is the normalization value at the flux decay phase,
the apparent area in the long burst in approximately two times larger (Fig.\,\ref{fig:HR}). 
The natural explanation for that is the presence of the optically thick  accretion disk which 
in the soft state blocks a significant part of the NS apparent surface, while the hot, optically thin, transparent accretion flow 
in the hard state does not affect much the   NS apparent area.
In addition to that effect, there could be additional differences in the physical conditions of the NS atmosphere 
(and thus in spectral hardening factors), as 
the boundary/spreading layer in the soft state forces the NS atmosphere to rotate close to Keplerian velocity \citep{IS99,SulP06}. 
We note that similar differences between short and long bursts were also found in another X-ray
bursting NS  \citep{GMH08, ZMA10}. 

\subsection{Distance to \source}

\label{sec:dist}

\source\ resides in the globular cluster Terzan 2. The distance to that
$D$=7.5$\pm$0.7 kpc was measured by 
\citet{1995AJ....109.1131K}, 
while \citet{OBB97} give  $D$=5.3$\pm$0.6 kpc, if $R_{\rm V}=A_V/{E(B-V)}=3.6$ (more suitable for red stars, 
see \citealt{GR95}), or  7.7$\pm$0.6 kpc if $R_{\rm V}=3.1$. 
To cover all possibilities, we assume further a flat distribution from 5.3 to
7.7 kpc with Gaussian tails of $1\sigma$=0.6 kpc on both ends.

\section{Method}
\subsection{Basic relations}

Here we briefly present some well known relations between observed and real physical NS parameters, 
which arise due to the gravitational redshift and the light bending, together with 
some other important equations. The observed luminosity $L_\infty$, effective temperature $T_\infty$ and
apparent NS radius $R_\infty$ are connected with the luminosity at the NS surface $L$, the effective temperature measured 
at the surface $T_{\rm eff}$, and the NS circumferential  radius $R$ and mass $M$
by the following relations \citep{LvPT93} 
\be \label{eq:ltr}
     L_{\infty} = \frac{L}{(1+z)^2},
     \quad T_{\infty}= \frac{T_{\rm eff}}{1+z},
      \quad R_{\infty}= R\ (1+z), 
\ee
with the redshift factor
\be
1+z=\left( {1-2GM/Rc^2} \right)^{-1/2}. 
\ee
The gravity $g$ on the NS surface is larger in comparison with the Newtonian case due to the general relativity effects
\be\label{eq:gdef}
  g=\frac{GM}{R^2}(1+z), 
\ee
therefore, the Eddington luminosity is larger too:
\be \label{eq:Ledd}
  L_{\rm Edd} = \frac{4\pi GMc}{\kappa_{\rm e}} \ (1+z)= 4\pi R^2 \sigma_{\rm SB}T_{\rm Edd}^4.
\ee
Here $T_{\rm Edd}$ is the maximum possible effective temperature on the NS surface, 
$\kappa_{\rm e}=0.2(1+X)$ cm$^2$ g$^{-1}$ 
is the electron (Thomson) scattering opacity, and $X$ is hydrogen mass fraction. 
The observed Eddington luminosity is smaller for higher~$z$ 
\be
\label{eq:Leddinf}
L_{\rm Edd,\infty} = \frac{4\pi GMc}{\kappa_{\rm e}}~\frac{1}{1+z}.
\ee
This is related to the observed Eddington flux 
\be
\label{eq:fedd}
F_{\rm Edd} =  \frac{L_{\rm Edd,\infty}}{4\pi D^2} = \frac{GMc}{\kappa_{\rm e}\,D^2}\ \frac{1}{1+z} 
\ee 
and the Eddington temperature 
\be \label{eq:tedd1}
T_{\rm Edd,\infty}  = \left(\frac{gc}{\sigma_{\rm SB} \kappa_{\rm e} }\right)^{1/4} \!\! \frac{1}{1+z} = \!\!
\left( \frac{F_{\rm Edd} }{\sigma_{\rm SB}} \right)^{1/4}  \!\! \left( \frac{R_{\infty}}{D} \right)^{-1/2} \!\! ,
\ee 
which is the effective temperature corresponding to the Eddington flux 
at the NS surface corrected for the gravitational redshift. 
We note here that the electron scattering opacity decreases with temperature \citep[e.g.][]{Pac83,PSS83} and 
at the typical temperature  of $\sim$3 keV in the upper layers of
the luminous NS atmospheres is reduced by about 7\%. 
This affects the luminosity where the Eddington limit actually is reached.

\subsection{Atmosphere models and color correction} 
  
Numerous computations of X-ray bursting NS atmospheres
\citep{London84,London86, Ebi87, Madej91, PSZ91, ZPS96,SulP06,SPW11}
show that their emergent spectra at high luminosities are close to diluted blackbody spectra due to strong energy exchange 
between high energy photons and 
relatively cold electrons at NS surface layers (Compton down-scattering)
\be
\label{eq:fbfit}
   F_E   \approx\ w\  B_E(T_{\rm c}=f_{\rm c} T_{\rm eff}), 
\ee  
where $f_{\rm c}$ is the color correction (or hardness) factor and $w$ is the dilution factor, which 
at high luminosities is very close to $1/f_{\rm c}^4$  \citep{SPW11}. 

Spectra observed from the X-ray bursting NSs are close to thermal and 
usually they are fitted by  a blackbody with two parameters: 
the observed color temperature $T_{\rm bb}$ and the normalization $K=(R_{\rm bb}\mbox{(km)}/D_{10})^2$.
It is easy to find the relations between various temperatures 
\be
\label{eq:tempdep}
    T_{\rm bb} = f_{\rm c}\ T_{\infty}= f_{\rm c}\ \frac{T_{\rm eff}}{1+z}= \frac{T_{\rm c}}{1+z}.
\ee
The observed blackbody flux is then 
\be
\label{eq:Fobs}
     F = \sigma_{\rm SB} T_{\rm bb} ^4  \frac{R_{\rm bb} ^2}{D^2}=\sigma_{\rm SB} T_{\infty}^4 \frac{R_{\infty}^2}{D^2}
\ee
and we can find the relation between the normalization and the NS radius
\be
\label{eq:rbbD2}
\frac{R_{\rm bb}^2}{D^2} =  \frac{R^2}{D^2} \frac{(1+z)^2}{f_{\rm c}^4} =  \frac{R_{\infty}^2}{D^2} \frac{1}{f_{\rm c}^4} .
\ee
These formulae can be transformed to the relation between color correction and normalization 
\citep{Penninx89,vP90}: 
\be \label{eq:KAfc}
K^{-1/4}=f_{\rm c} A,  \qquad A=(R_{\infty} [{\rm km}]/D_{10})^{-1/2}.
\ee
A combination of  $A$ and $F_{\rm Edd}$  gives the Eddington temperature:  
\be \label{eq:tedd}
T_{\rm Edd,\infty}  = 1.14\times 10^8\, A\,F_{\rm Edd, -7}^{1/4}\ \mbox{K}
= 9.81\, A\, F_{\rm Edd, -7}^{1/4}\ \mbox{keV} , 
\ee 
where $F_{\rm Edd, -7} = F_{\rm Edd} / 10^{-7}$ erg cm$^{-2}$ s$^{-1}$.

A detailed comparison of the theoretical models with the data requires  the knowledge of the run of the 
color correction with flux. Previous models covered the range of luminosities very sparsely. 
Using our recently developed code \citep{SulP06,SMD06,SW07}, 
we have computed  a very detailed set of models with the luminosity varying by three orders of magnitude \citep{SPW11}.

\begin{figure}
\plotone{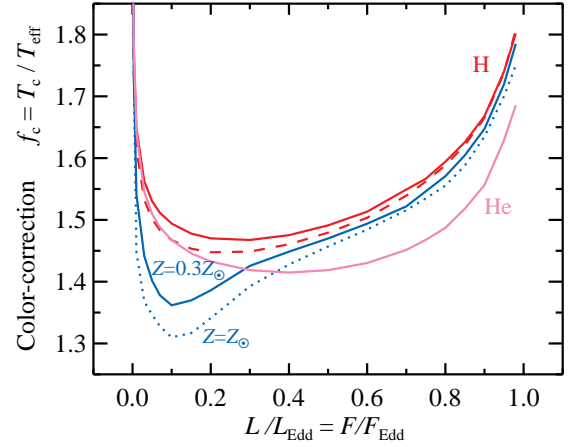}
%\centering 
%\includegraphics[width=0.95\columnwidth]{fig6.eps}
\caption{Color-correction factor as a function of the NS luminosity \citep{SPW11}. 
The curves correspond to atmospheres of different chemical composition:
pure hydrogen (red), pure helium (pink); the blue curves are for models
with solar H/He composition plus solar abundance of metals $Z=Z_{\odot}$ 
(dotted blue) and subsolar metals $Z=0.3Z_{\odot}$ (solid blue). 
The surface gravity is taken to be $g = 10^{14.0}$ cm s$^{-2}$. 
The dashed curve shows the results for a hydrogen atmosphere at larger gravity of $\log g = 14.3$.}	
\label{fig:fc}
\end{figure}

An atmosphere model is fully defined by the surface gravity $g$,
chemical composition, and the ratio of the luminosity to the Eddington luminosity $l=L/L_{\rm Edd}$. The last parameter also 
relates the effective temperature to the Eddington temperature at the NS surface: 
\be \label{eq:teffl}
   T_{\rm eff} = l^{1/4} T_{\rm Edd}.
\ee
We considered various chemical compositions (pure hydrogen, pure helium, solar
mixture of hydrogen and helium with 
various metal abundances) and three surface gravities $\log g=14$, 14.3, and
14.6. The (redshifted) radiation spectrum 
from the NS atmosphere 
was then fitted with  a diluted Planck function in the 3--20 keV energy band
(i.e. the range observed by \rxte) 
to determine the color-correction factor $f_{\rm c}$ (see Fig. \ref{fig:fc}).  
The behaviour of $f_{\rm c}$ at relatively high $l$ depends mainly on the hydrogen abundance $X$ and very little on 
the surface gravity and metal abundance.

\subsection{Determining $M$ and $R$ using the touchdown method}
\label{sec:mr_fedd_K}

In an ideal situation if the observed X-ray emission  comes indeed from the passively cooling fully visible NS 
and the distance to the source is known, we can determine NS mass $M$ and radius $R$ from two observables: 
the Eddington flux given by Equation (\ref{eq:fedd}) and the NS apparent
blackbody size in the cooling tail, 
or quantity $A=K^{-1/4}/f_{\rm c}$ 
\citep[see e.g.][]{LvPT93}. 
The latter is related to the apparent size of the NS through the color correction, $R_{\infty}=f_{\rm c}^2 R_{\rm bb}$, 
and $f_{\rm c}$ is assumed to be known in the burst tail from the theoretical considerations. 
While this method was proposed long time ago, 
only recently strong claims appeared in the literature that it actually can be used for 
determining accurately parameters of three bursting NSs \citep{OGP09,GO10, GWCO10}. 

\begin{figure}
\plotone{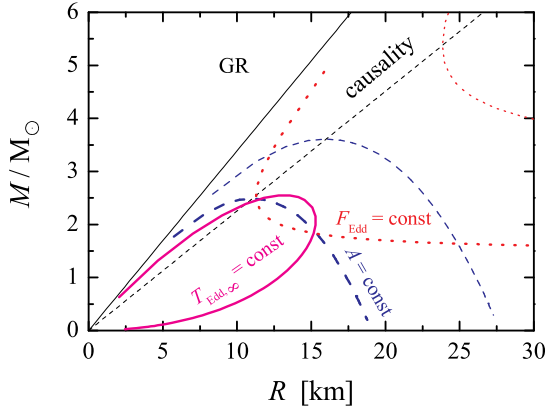}
%\centering 
%\includegraphics[width=0.95\columnwidth]{fig7.eps}
\caption{Constraints on $M$ and $R$ from various observed values. 
The solid curve gives the relation obtained from the Eddington temperature given by Equation (\ref{eq:tedd}), 
thick dotted curve is for the Eddington flux given by Equation (\ref{eq:fedd}), thick dashed curve is for $A$=const. 
If the assumed distance is too large, there are no solutions 
(the corresponding curves for $F_{\rm Edd}$=const and $A$=const shown by thin lines do not cross).   }	
\label{fig:curv}
\end{figure}

Using the approach advertised in the aforementioned papers, one has to determine 
the Eddington flux from observations. For PRE bursts it was assumed that it is reached at the ``touchdown'' point
\citep{Damen90}, when the color temperature is highest and the apparent blackbody area lowest.  
The color correction factor $f_{\rm c}$ at the late cooling phases of the PRE
bursts was  taken to be close to 1.4 \citep{OGP09,GO10, GWCO10} 
 based of the models by \citet{MJR04} and \citet{MM05} (see discussion below).   
The observables can be then transferred to the constraints on 
$M$ and $R$ (see Fig. \ref{fig:curv}). We will further call this approach the ``touchdown method''.

From the Eddington flux estimate we have (see dotted curves in Fig.\,\ref{fig:curv})
\bea \label{eq:eddflux}
R & =&  \frac {2 \kappa_{\rm e}  D^2 F_{\rm Edd} }{c^3 } \ u^{-1}\ (1-u)^{-1/2}  \nonumber \\
&=& 14.138\  {\rm km}\ (1+X)\ D_{10}^2\ F_{\rm Edd, -7}  \ u^{-1}\ (1-u)^{-1/2} 
\eea
 and  the mass is found using 
 \be \label{eq:m_r_u}
\frac{M}{\msun} =  \frac{R}{2.95\ {\rm km} } \ u ,
\ee
where the compactness $u=R_{\rm S}/R=1-(1+z)^{-2}$ and $R_{\rm S}=2GM/c^2$ is the Schwarzschild radius of the NS.
A measurement of $A$ gives another  constraint: 
\be \label{eq:ra}
R = R_{\infty} \sqrt{1-u} = f_{\rm c}^2 \ D \sqrt{K}  = D_{10}\ A^{-2} \ \sqrt{1-u}\ \mbox{km} . 
\ee 
Combining with the parametric expression (\ref{eq:m_r_u}) for the mass, we get the second  
relation between $M$ and $R$ shown by the dashed curves in Fig.\,\ref{fig:curv}.

The Eddington temperature  given by Equation (\ref{eq:tedd})
 is independent of the uncertain distance to the source and
can be used to express the NS radius through the observables and  $u$: 
\be \label{eq:tedd2}
R =  \frac{c^3}{2  \kappa_{\rm e}  \sigma_{\rm SB} T_{\rm Edd,\infty}^4 } \ u\ (1-u)^{3/2} ,
\ee
and the mass is then found via Equation (\ref{eq:m_r_u}). The corresponding 
relation between $M$ and $R$ is shown by the solid curve in Fig.\,\ref{fig:curv}. 

All three curves cross in one or two points (see Fig.\,\ref{fig:curv}) 
if the quadratic equation 
\be
u(1-u) = \frac{2\, \kappa_{\rm e}\ D \ F_{\rm Edd}}{\sqrt{K}\ f_{\rm c}^2\ c^3}  = 14.138\,(1+X)\,D_{10}\,F_{\rm Edd, -7}\,A^2 , 
\ee  
which follows from Equations (\ref{eq:eddflux}) and (\ref{eq:ra}), has a real solution for $u$
\citep[see e.g.][]{SLB10}. This happens if $u(1-u)<1/4$ and 
the distance then should satisfy the following inequality 
\be \label{eq:distmax}
D \leq D_{\max} = \frac{\sqrt{K} f_{\rm c}^2 c^3}{8\,F_{\rm Edd}\,\kappa_{\rm e}} = 
\frac{0.177}{(1+X)\ A^2\ F_{\rm Edd, -7} } \ \mbox{kpc} .
\ee
In the opposite case, there is no physical solution for $M$ and $R$ for given observables.

As we mentioned above, this method of determination of $M$ and $R$ works in an ideal situation. 
There are a few problems with this approach. 
First, the relation of the Eddington flux to the touchdown flux is not clear. 
 The reduction of the electron scattering opacity  at high temperatures 
increases the true Eddington limit by about 7\% above that given by Equations (\ref{eq:Leddinf}) and (\ref{eq:fedd}).  
Also if we believe that the X-ray burst luminosity is equal to the Eddington luminosity during the expansion phase of the PRE burst, 
the observed luminosity has to decrease when the photospheric radius decreases
according to Equation (\ref{eq:Leddinf}) \citep{LvPT93}.
In reality the observed luminosity in \source\ increases when the photospheric radius decreases (see Fig. \ref{fig:obs}). 
This implies that the ratio of the luminosity to the Eddington limit at the
photosphere has to increase with  decreasing photospheric radius (i.e. increasing redshift $z$). 
A combination of this dependence with the gravitational redshift effect predicts then that 
the observed luminosity reaches the maximum when the photospheric radius is larger than $R$ and  
that maximum is larger than expected for the Eddington luminosity at the surface.

Second, the assumption of $f_{\rm c}\approx 1.4$ in the cooling tail is very uncertain. 
This assumption is based on X-ray burst atmosphere models of \citet{MM05}, who claimed a rather constant $f_{\rm c}$ at 
low effective temperatures (see also Fig.\,6 in \citealt{GO10}), as well 
as the fact that most of the short PRE bursts have a constant 
normalization at   late phases. 
We note here that the factors $f_{\rm c}$ in \citet{MM05} correspond to the
ratio of the energy where 
the peak of the model flux $F_{E}$
is reached to the peak energy of the blackbody spectrum at effective
temperature. Moreover, the low-luminosity 
models were calculated 
for high surface gravity instead of low effective temperatures, which leads to incorrect results \citep[see][ for details]{SPW11}.
The color-corrections obtained in this way, however, should not be compared to the data at all, 
because the color temperatures of the time-resolved spectra from 
X-ray bursts are computed by {\it fitting} the actual data in a specific energy interval (e.g. 3--20 keV for \rxte/PCA)  
with the diluted blackbody function with arbitrary normalization. The values of $f_{\rm c}$ shown in Fig. \ref{fig:fc} on the other 
hand are produced using the procedure similar to that applied to the data, i.e. by fitting the model spectra in the 3--20 keV range.
As was shown by \citet{SPW11} the color-correction has a 
flat part at $l\sim 0.2$--0.5 for most of the chemical compositions, but the actual value of $f_{\rm c}$ depends 
on the hydrogen fraction, and, for example, for $X=1$ it is closer to 1.5 than to 1.4 (see Fig. \ref{fig:fc}). At lower luminosities 
$f_{\rm c}$ can first drop because of iron edges and then increase to rather high values. Thus, 
there is no unique value for $f_{\rm c}$ in the cooling tail. 
The expected significant variations of $f_{\rm c}$ with flux also imply that constancy of the 
apparent blackbody area in the cooling tail contradicts the burst atmosphere models and therefore 
the bursts showing such behaviour  obviously demonstrate influence of some
physics not included in these simplest models 
of NS atmospheres, and thus cannot be used for determination of NS masses and radii with the help of the aforementioned models.

Third, different PRE bursts from the same source show different cooling tracks, for example, 
the long burst and the short bursts of \source\ (see  Fig.\,\ref{fig:obs}) have normalizations different by a factor of two. 
This makes the determination of the apparent area from a single burst not unique.

And finally, the most serious problem with this approach is that out of the whole amount of information on 
the cooling tail of the burst, one uses only {\it two} numbers and it is not checked whether these quantities 
are actually consistent with each other. For example, the theory also predicts that the 
color  correction $f_{\rm c}$ changes from  $\approx$1.7 to $\approx$1.4 when the luminosity drops from 
the Eddington to about 1/3 of the peak value. This also implies  that the blackbody normalization between the touchdown point 
and the decay phase must increase at least  by a factor of two.  It is really true for the long burst from \source,
while the short bursts have nearly constant $K$, which is two times smaller than that in the long burst, implying 
probably a partial eclipse of the NS by the optically thick accretion disk and/or 
the influence of the boundary layer on the structure of the NS atmosphere as discussed in Section \ref{sec:data}. 
We also note here that all bursts analyzed   by \citet{OGP09} and \citet{GO10,
  GWCO10} are short, they do not 
show enough variations of $K$ 
in their cooling tracks, and therefore the results obtained from these bursts
are not reliable 
(see Sect. \ref{sec:ozel} for more details).

On the base of all these arguments we offer a new approach to the NS mass and radius estimations using 
the information from the whole cooling track.

\subsection{Determining $M$ and $R$ using the cooling tail method}
\label{sec:mr_cooling}

\begin{figure}
\plotone{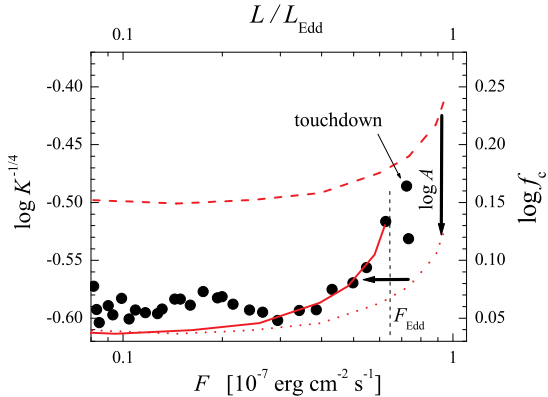}
%\centering 
%\includegraphics[width=0.99\columnwidth]{fig8.eps}
\caption{Illustration of the suggested new cooling tail method. 
The dependence $K^{-1/4}$--$F$ as observed during the cooling track of the long burst from 
4U\,1724--307 on November 8, 1996 (circles). 
The theoretical $f_{\rm c}$--$L/L_{\rm Edd}$ dependence is shown by the dashed curve (right and upper axes) 
and the best-fit relation (solid curve). }	
\label{fig:meth}
\end{figure}

If  the radiating surface area does not change during the burst decay phase, 
the evolution of the normalization is fully determined by the color correction variations (see Equations\,(\ref{eq:rbbD2}) and (\ref{eq:KAfc})). 
We thus suggest to fit the observed relation $K^{-1/4}$--$F$ at the cooling phase of the burst by the 
theoretical relations $f_{\rm c}$ -- $L/L_{\rm Edd}$  (shown in Fig. \ref{fig:fc}, see also \citealt{SPW11}) 
with free parameters being $A$  and the Eddington flux $F_{\rm Edd}$ (see Fig.\,\ref{fig:meth} for illustration). 
The behaviour of $f_{\rm c}$ depends rather weakly  on the NS gravity and chemical  composition, 
which substantially reduces the model dependence of the fitting procedure.   
Using the obtained best-fit parameters, we can then apply the method identical to that
described in Section \ref{sec:mr_fedd_K}.

The main advantages of the proposed  cooling tail method 
is that there is no freedom in choosing $f_{\rm c}$ in the cooling tail, 
the determination of the Eddington flux becomes decoupled from the uncertainties related to the touchdown  flux as 
the whole cooling tail is used, and finally, one can immediately check whether 
the burst spectral evolution is consistent with theoretical models 
and whether the employed model includes the majority of the 
relevant physics for the description of the considered phenomenon.
This check can help to choose for further analysis only those bursts that follow the theory.

\section{Results} 
 
\subsection{The long burst from \source}

\subsubsection{Determining NS parameters using the cooling tail method}
 
\begin{figure}
\plotone{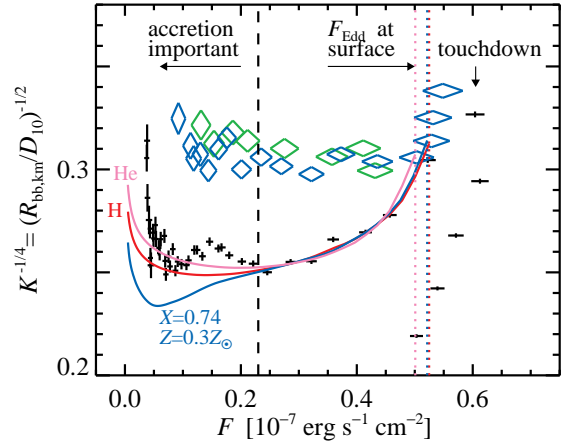}
%\centering 
%\includegraphics[width=0.95\columnwidth]{fig9.eps}
\caption{Comparison of the X-ray burst data for \source\ to the theoretical models of NS atmosphere.
The crosses present the observed dependence of $K^{-1/4}$ vs. $F$ for the long burst, while diamonds represent two short bursts for 
the blackbody model with constant absorption $N_{\rm H}=10^{22}$ cm$^{-2}$.
The solid curves correspond to the three best-fit theoretical models 
of various chemical compositions (see Fig. \ref{fig:fc}). 
The best-fit parameters $F_{\rm Edd}$  and $A$, defined by Equations  (\ref{eq:fedd}) and (\ref{eq:KAfc}), are given in Table~\ref{tab:fit}.  }	
\label{fig:data}
\end{figure}

Let us apply the method described in Section \ref{sec:mr_cooling} for determining 
NS mass and radius from the data on the long burst from \source.
We fit the dependence of the normalization constant $K$ on the observed flux $F$  for the long burst by
the theoretical curves $f_{\rm c}$ -- $L/L_{\rm Edd}$ computed for three chemical compositions. 
They give a good description of the data at  intermediate fluxes
for the data points to the right of the vertical dashed line (Fig. \ref{fig:data}), but below the touchdown, which 
we use for fitting. 
Close to the touchdown, significant deviations are probably caused by deviations from the plane-parallel atmosphere
and effects of the wind (thus the models are not reliable). Strong deviations are also visible at low fluxes where 
the burst spectrum is probably modified by accretion. 
Assuming that $F_{\rm Edd}$ is actually reached at the touchdown contradicts the following evolution of the parameters during the cooling phase. 
The fits are better for  the hydrogen-rich atmospheres. 
The results of the fitting for all considered chemical compositions of the NS atmosphere are presented in Table\,\ref{tab:fit}.
The uncertainties in $A$ and $F_{\rm Edd}$ are obtained with a bootstrap method.

\begin{deluxetable*}{lccccccc}
\tabletypesize{\scriptsize}
\tablecaption{Best Fit Parameters \label{tab:fit}}
\tablewidth{0pt}
\tablehead{
\colhead{Atmosphere model} & \colhead{$F_{\rm Edd}$} & \colhead{$A$}  & \colhead{$T_{\rm Edd,\infty}$} & 
\colhead{$\chi^2$/d.o.f.} & \colhead{$M$} & \colhead{$R$}  \\
 &  (10$^{-7}$ erg s$^{-1}$ cm$^{-2}$) &  (km/10 kpc)$^{-1/2}$ & ($10^7$ K) &     & $(\msun)$ &  (km)  }
\startdata
Hydrogen                                 & 0.525$\pm$0.025 & 0.170$\pm$0.001 & 1.64$\pm$0.02 &  5.0/5  &  1.9$\pm$0.4 / 2.45$\pm$0.15    & 14.7$\pm$0.8  /  11.7$\pm$1.3      \\
                                  &  &  & &  &  1.4 (fixed)    & 14.2$\pm$0.4       \\
Solar H/He, $Z=0.3Z_{\odot}$ & 0.521$\pm$0.020 & 0.172$\pm$0.002 & 1.66$\pm$0.02 &  5.8/5  &   1.85$\pm$0.6 /   2.7$\pm$0.15    &  15.5$\pm$1.5 /  13.0$\pm$1.0      \\
                                  &  &  & &  &  1.4 (fixed)    & 15.2$\pm$0.4       \\
Helium   			       & 0.50$\pm$0.02     & 0.178$\pm$0.002 & 1.71$\pm$0.02 &  11.3/5 &   1.05$^{+0.55}_{-0.4}$ & 18.0$^{+3.5}_{-3.5}$   \\
                                  &  &  & &  &  1.4 (fixed)    & 20.2$\pm$0.5       
 \enddata
\tablecomments{Results of the fits to the $K^{-1/4}$--$F$ dependence with the 
NS atmosphere models  for various chemical compositions  and $\log g=14.0$.  
For hydrogen and solar composition atmospheres there 
are two solutions for $M$ and $R$ (see Fig. \ref{fig:mr}).
Neutron star mass and radius are computed from $A$ and $F_{\rm Edd}$ assuming 
a flat distribution of the distance between 5.3 and 7.7 kpc with  Gaussian tails of 1$\sigma$=0.6 kpc.
Errors correspond to the 90\% confidence level.  }
\end{deluxetable*}

\begin{figure}
\plotone{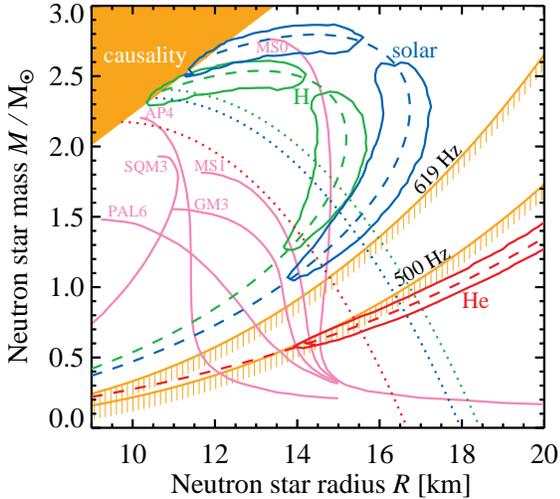}
%\centering 
%\includegraphics[width=0.8\columnwidth]{fig10.eps}
\caption{Constraints on the mass and radius of the NS in  \source\ from the long burst spectra (fitted with the blackbody model
and constant absorption). 
The dotted curves correspond to the best-fit parameter $A$ for the distance to the source of 5.3 kpc.  
For a flat distribution of the distance between 5.3 and 7.7 kpc with  Gaussian tails of 1$\sigma$=0.6 kpc, 
the constraints are shown by contours (90\% confidence level). 
They are elongated along the (dashed) curves  corresponding to the Eddington temperatures $T_{\rm Edd,\infty}$ 
given by Equation (\ref{eq:tedd}) (which do not depend on the distance). 
These correspond to the three chemical compositions: green for pure hydrogen,  blue for the solar ratio of H/He and subsolar metal abundance $Z=0.3Z_{\odot}$ appropriate for Terzan 2 \citep{OBB97}, and red for pure helium. 
The mass-radius relations for several equations of state of neutron and strange stars matter are shown by solid pink curves. 
The upper-left region is excluded  by constraints from the causality requirements \citep{HPY07,LP07}.
The brown solid curves in the lower-right region correspond to the mass-shedding limit and delineate the zone forbidden for \source, if it had a rotational frequency of 500 or 619 Hz, the highest detected for the X-ray bursters \citep{SB06}.
%The hatched horizontal belt marks the spread of pulsars masses accurately measured in double NS binaries \citep{HPY07}. 
}	
\label{fig:mr}
\end{figure}

Taking the distance  in the range 5.3--7.7 kpc (see Section \ref{sec:dist}), 
we convert a distribution of $F_{\rm Edd}$  and $A$ using Monte-Carlo simulations 
to the distribution of $M$ and $R$ (Fig. \ref{fig:mr} and Table~\ref{tab:fit}). 
The resulting contours are elongated, because of the uncertainty in distance,
along the curves of constant Eddington temperature. 
 The pure helium model atmospheres give a mass which is too small from the stellar evolution point of view. 
It is also below the mass-shedding limit if the star is rotating faster than at about 500 Hz. 
Pure hydrogen atmosphere models are consistent with the data only for $D<6$
kpc, while for the atmosphere of 
solar composition the upper limit is 7 kpc.  The hydrogen rich atmosphere
models give a lower limit on the 
stellar radius of 14 km independently of the metal abundance (see
Table~\ref{tab:fit}) for NS masses less than 2.3$\msun$, 
and smaller radii are allowed only for high NS masses.  
 For the helium atmosphere, the solution shifts towards higher masses and
 larger radii exceeding the mass-shedding limit 
(e.g.  for a 1.5 solar mass star the radius is about 20 km).
If we take the canonical neutron star mass of 1.4$\msun$,
the NS radius is then strongly constrained at $R=15.2\pm0.4$ km assuming  solar H/He composition with  $Z=0.3Z_{\odot}$
and 14.2$\pm0.4$ km for hydrogen. 
The obtained constraints (see Table~\ref{tab:fit}) imply a stiff equation of state of the NS matter. 

The choice of the highest-flux point at the $K^{-1/4}$--$F$ plot used in the fitting procedure affects the results slightly. 
Neglecting even the second point after the touchdown, 
reduces the estimated $F_{\rm Edd}$  by 3\%, while $A$ (defined by the horizontal part of the cooling track)
remains unchanged. This leads to a 2\% increase in radius and 4\% decrease in the NS mass estimates. 
The fits to the  $K^{-1/4}$--$F$ dependence, which is obtained with blackbody fits with free $N_{\rm H}$, 
in the same flux intervals as for constant $N_{\rm H}$ 
give values of $A$ about 10\% smaller, while the Eddington flux estimates remain the same within 1\%. In this case, 
the estimated NS radii grow by 20\% relative to those shown in Fig.~\ref{fig:mr}. 

 In addition, there is a systematic uncertainty of about 10\% in the absolute values of fluxes measured by the \rxte\ \citep{Kirsch05,Weisskopf10}.
It acts similarly to an additional 5\% uncertainty on the distance and does not affect the value 
of $T_{\rm Edd,\infty}$. With the current uncertainty on the distance to \source, 
this additional inaccuracy does not substantially increase the error bars on $M$ and $R$  in Table~\ref{tab:fit}.

The determined Eddington flux is smaller than the touchdown flux by about 15\%. 
The main part of this difference can be easily explained by the
temperature dependence of the electron scattering opacity \citep{Pac83,PSS83,LvPT93}.
In our model calculations \citep{SPW11}, we used the Kompaneets equation 
to describe the electron scattering and this assumes the Thomson opacity. 
As the upper atmosphere layers can be as hot as 3--3.5 keV, 
the electron scattering opacity there is smaller than the Thomson one by about 6--8\% and
the actual Eddington limit is reached at correspondently higher luminosity than $L_{\rm Edd}$ given by Equation (\ref{eq:Leddinf}). 
Because we fit the data at luminosities much below the Eddington (where correction to the Thomson opacity are small), 
we determine $F_{\rm Edd}$ as given by Equation (\ref{eq:fedd}), which is used for determination of $M$ and $R$.
We also note here that the opacity effect is not included in the touchdown method, which 
assumes the touchdown flux being equal to the Eddington flux at Thomson opacity. 

The remaining $\sim$8\% difference between the ``true'' Eddington flux and the touchdown flux 
can be understood, if we take into account that the maximum luminosity for PRE bursts can exceed 
the Eddington luminosity on the surface  due to the dependence of the observed Eddington luminosity on the
redshift $z$ (see Equation\,(\ref{eq:Leddinf}) and \citealt{LvPT93}). This would imply that 
the photospheric radius corresponding to the touchdown exceeds the NS radius by $\sim$25\%. 
The corresponding color correction then has to be $\sim$15\% larger than for our models  with $l=0.98$, 
which is consistent with that expected at $L\sim L_{\rm Edd}$ \citep{PSZ91}. 
The sharp maximum in $T_{\rm bb}$ (and minimum in $K$) can arise from 
a joint influence of the increasing color correction and decreasing effective temperature 
during the photospheric radius expansion phases.

The lower limit on the NS radius of 13.5--14 km as obtained by us is consistent 
with the measurements for the thermally emitting quiescent NS X7 in the globular cluster 47 Tuc
($14.5^{+1.8}_{-1.6}$ km; \citealt{HR06}). 
However, the radii of other thermally emitting quiescent NSs are significantly smaller (9--13 km; \citealt{WB07,GRB11}).
We note that these results depend on the model of the NS hydrogen 
atmosphere, which were also computed for the passively cooling NSs  \citep{HR06}. 
In case a quasi-spherical accretion occurs at a low rate  during the quiescence, the additional heat 
dissipated in the upper atmosphere would increase the temperature there, and, therefore, ~$f_{\rm c}$ \citep[see e.g.][]{ZTT00}. 
In this case, the NS radius can be underestimated. 
We also need to note that if the NS in \source\ has a spin of 500 Hz, 
the radius of the non-rotating star would be about 1 km smaller than estimated above. 
Our constraints on the NS radii are in good agreement with the NS radii (13--16 km) 
evaluated by \citet{SulP06} from the spectra of low-mass X-ray binaries using the spreading layer model. 
Rather large NS radii are allowed by the modern equations of state \citep{HLP10}, 
which predict the upper limit of 13.5 km for a 1.4$\msun$ NS.

\subsubsection{Touchdown method}

\begin{figure}
\plotone{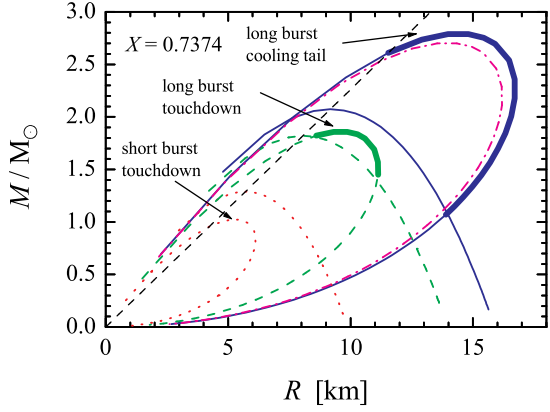}
%\centering 
%\includegraphics[width=0.95\columnwidth]{fig11.eps}
\caption{Comparison of the solutions (thick curves) obtained for the long burst using the cooling tail method 
(solid curves) and the touchdown method (dashed curves), 
and for the short bursts using the touchdown method only (dotted curves). 
The curves corresponding to the obtained $T_{\rm Edd,\infty}$ and $A$ 
(at $D_{10}=0.47$) are also shown. All solutions are derived for $X$=0.7374.
The dash-dotted curve corresponds to the short burst with $K=115$, $F_{\rm Edd, -7} = 0.55$ and 
$f_{\rm c}=1.78$ (i.e. $T_{\rm Edd,\infty} = 1.45$~keV).}	
\label{fig:solsh}
\end{figure}

We apply now the touchdown method  described in Section \ref{sec:mr_fedd_K} 
to the same data on the long burst.  
This approach is used for illustrative purpose only
and we do not consider here any statistical errors.
Let us take the Eddington flux equal to the touchdown flux $F_{\rm Edd, -7}$=0.605, 
the normalization in the cooling tail $K$=230,  $f_{\rm c}$=1.4 and $X$=0.7374. 
From these quantities we obtain $T_{\rm Edd,\infty} = 1.84\times10^7$ K 
and an upper limit on the distance $D_{10, \max} =0.5$. 
The curves relevant to the derived $T_{\rm Edd,\infty}$ 
and the distance $D_{10}$ = 0.47 (corresponding to 1$\sigma$ deviation from the minimal distance of 5.3 kpc) 
are shown in Fig.\,\ref{fig:solsh} by dashed lines. 
Using this approach  we can estimate $M \approx (1.6\pm 0.2) \msun$ and $R \approx 10.0\pm1.5$ km. 
We see that the touchdown method gives a substantially smaller NS radius compared to the cooling tail method.

\begin{figure}
\plotone{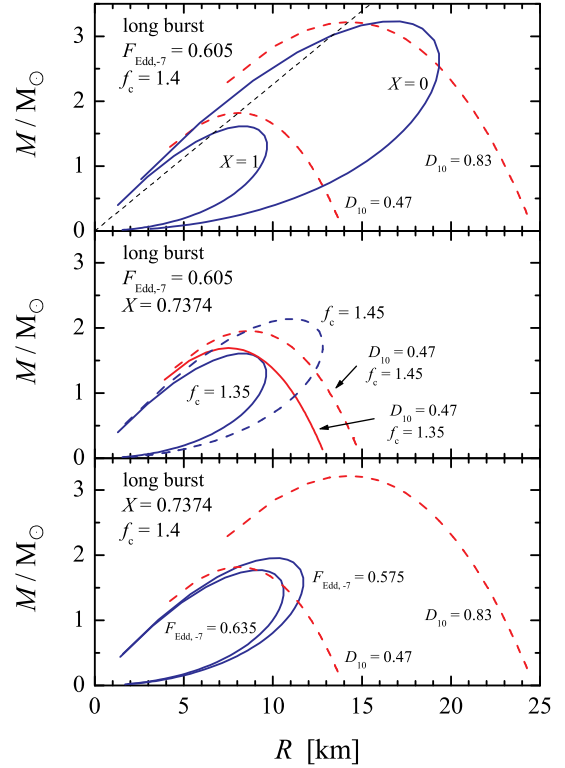}
%\centering 
%\includegraphics[width=0.9\columnwidth]{fig12.eps}
\caption{Expected uncertainties for the $M$ and $R$ solutions obtained for the long burst using the touchdown method.}	
\label{fig:uncrt}
\end{figure}

The uncertainties in $M$ and $R$ are actually much larger, because of the uncertainties 
in chemical composition, color-correction and distance (see  Fig.\,\ref{fig:uncrt}).
The most significant errors arise due to unknown distance and the 
hydrogen mass fraction in the NS atmosphere (see top panel of Fig.\,\ref{fig:uncrt}, where 
we considered two limiting cases for the  distance of $D_{10}=0.47$ and 0.83 corresponding to 1$\sigma$ deviations 
on both ends of the distance distribution and for the hydrogen mass fraction $X=1$ and 0). 
Uncertainties in  $f_{\rm c}$ are also affecting the results  
(middle panel of Fig.\,\ref{fig:uncrt}), for example, changing $f_{\rm c}$ from 1.35 to 1.45 increases the 
maximum possible $R$ from 10 to 13 km and $M$ from 1.5 to 2$\msun$. 
We note that $f_{\rm c}$ is actually closer to 1.5 for the hydrogen-rich atmospheres (see Fig.\,\ref{fig:fc}).     
Less significant errors appear due to uncertainties in the observed Eddington flux on the NS surface (bottom panel of Fig.\,\ref{fig:uncrt}). A larger flux corresponds to smaller $M$ and $R$.

\subsection{Short bursts}

The cooling tracks for the short PRE bursts shown in Fig. \ref{fig:data} are very different from that of the long burst 
and are completely inconsistent with the theoretical dependences. This is a strong argument that these cooling tracks are 
affected by some additional physics and cannot be used for  determination of NS mass and radius. 
Ignoring that fact, let us still apply the touchdown method to the short bursts. 
Taking from the observed data (see Fig.~\ref{fig:obs}) 
 $F_{\rm Edd,-7} \approx 0.55$ and $K\approx 115$ and assuming $f_{\rm c}=1.4$, we find that 
 the curves on the $M$--$R$ plane corresponding separately to constraints from $F_{\rm Edd}$ and $K$ do not cross 
for hydrogen-rich atmospheres ($X>0.7374$) at any distance  larger than 3.8 kpc, while 
for $D=4.7$ kpc they cross only for $X<0.45$ (see Figs.\,\ref{fig:solsh} and \ref{fig:data}).  
The situation that the two observables are consistent with each other in a very restricted range of $X$ 
and distances is not unique to \source, but typical for many bursters, 
particularly those analyzed by \citet{OGP09} and \citet{GO10} as was shown by \citet{SLB10}, see Section \ref{sec:ozel} for details. 
For pure He atmospheres, there are consistent solutions at $M\sim 1.5\msun$ and $R\sim 10$ km at the distance $\sim 6$ kpc.  
However, we note again that the results from these short bursts are not reliable, because their cooling tracks 
contradict the theory the analysis is based on (which predicts $f_{\rm c}\sim 1.4$ in the cooling tail and much higher at 
luminosities close to the Eddington).

The measured $K \approx 115$  from the horizontal part of Fig.\,\ref{fig:obs} bottom panel  
is two times smaller than corresponding values for the long burst $K \approx 230$. 
This can result from two effects simultaneously.
The long X-ray burst occurred in the hard state, when the accretion flow in the NS vicinity had a small optical depth which 
only marginally affects the NS  photosphere and it cannot eclipse the neutron star.  On the other hand, the short bursts happened  
in the soft state, when the persistent emission originated in the optically thick accretion disk and the boundary layer.
Therefore, the accretion disk can just block part of the star in the decay phase of the burst 
reducing $K$ by factor of up to two for the case of large inclination. 
Even if the inclination is small, the apparent NS area decreases by a factor of  $\kappa=1-u^2$ (where $u=R_{\rm S}/R$). 
These limiting cases can be united in a simple approximate formula for the reduction factor 
\be
\kappa = \frac{1}{2} \left( 1+ \left[1-2u^2\right] \cos i \right) . 
\ee
An additional effect can be related to the optically thick boundary layer. 
If the spreading layer model describes the boundary layers correctly \citep{IS99},    
a significant part of the emergent radiation can arise in the rapidly rotating spreading layer above the hot NS surface, 
which has a reduced effective gravity due to the  centrifugal force, resulting in a flux through the atmosphere close to the local 
Eddington limit and a high color correction $f_{\rm c}\approx$ 1.6--1.8 \citep[see ][]{SulP06} and therefore small $K$.
Using $K^{-1/4} \approx 0.305$ as measured in the short bursts  and $A\approx0.172$ as determined from the long burst, 
we can estimate then the color-correction at the  cooling tails of the short bursts $f_{\rm c}\approx1.77\,\kappa^{1/4}$, 
which accounts for both effects (see Fig.\,\ref{fig:solsh}). 
The importance of the spreading layer can vary with the accretion rate 
and potentially the normalization values from $K$ for the long burst to $K/2$ can be found  \citep[see e.g.][]{ZMA10}.

It is also possible that the chemical composition of the NS atmosphere during short and long bursts is different. For example,
pure helium atmospheres give $K$ by 20--25\% larger than pure hydrogen atmospheres because $f_{\rm c}$ for helium atmospheres is 
5--6 \% smaller (see Equation\,(\ref{eq:rbbD2}) and Fig.\,\ref{fig:fc}). 
However, the maximum temperature at the touchdown point must be  $\approx$ 20\% larger for
helium atmospheres ($\approx 3.2$ keV, see Equation\,(\ref{eq:tedd})), while 
the maximum temperatures in the long and the short bursts are very 
close to each other, $\approx$ 2.7 -- 2.8 keV, which argues against a difference in the 
chemical composition. 

\section{Comparison to other X-ray bursters}
\label{sec:other}

\subsection{Long PRE bursts}

\citet{Penninx89} analyzed two long-duration ($>$100 s) PRE bursts observed by {\it EXOSAT}/ME in 1984 and 1986 
from \mbox{4U\,1608--52} in its hard state at a rather low persistent flux of
(1--2)$\times 10^{-9}$ erg cm$^{-2}$ 
s$^{-1}$ in the 2--20 keV band. 
The evolution of $T_{\rm bb} F^{-1/4}$ (which is proportional to $K^{-1/4}$) with $F$ shown in their Fig.\,7
is almost identical to the models presented in Fig.\,6 of \citet{SPW11}. 

\citet{Kuulkers03} reported observations by \rxte/PCA  of the long PRE bursts from \source\ (analyzed here), 
the atoll source \mbox{4U\,2129+11} in globular cluster M15, and \mbox{H1825--331} in globular cluster \mbox{NGC\,6652}. 
Spectral evolution after the touchdown in all sources is very similar. 
\citet{Kuulkers03} also reported observations by {\it BeppoSAX}/WFC of 24 long PRE bursts from \source. 
Spectral evolution during the cooling phases of these bursts is entirely consistent with that observed 
in the long burst by \rxte/PCA. 

Observations by {\it Ginga}/LAC of a long PRE burst from \mbox{4U\,2129+11} during  its island (hard) state 
at a low persistent flux level of $\sim0.5\times 10^{-9}$ erg cm$^{-2}$ s$^{-1}$ are presented by \citet{vP90}.
The behaviour of   $T_{\rm bb} F^{-1/4}$  at fluxes above 30\% of the peak (touchdown) flux 
shown in their Fig. 10 is very similar to that for the long burst from \source\ shown in our Fig. \ref{fig:data}. 
For both objects the data at high fluxes are well described  by the theory. We note that in both cases 
the position of the touchdown point in the $K^{-1/4}$--$F$ diagram is not
consistent with extrapolation of the data 
from intermediate fluxes, 
implying that the Eddington flux is smaller  than the touchdown flux.

\subsection{Short PRE bursts from \mbox{EXO\,1745--248}, \mbox{4U\,1820--30} and \mbox{4U\,1608--52}}
\label{sec:ozel}

Recently strong claims appeared in the literature that both the NS mass and radius can be determined 
with an accuracy of better than 10\% from the PRE bursts from three NSs: 
\mbox{EXO\,1745--248} ($M=1.7 \pm 0.1 \msun, R=9\pm 1$ km, \citealt{OGP09}), 
\mbox{4U\,1608--52} ($M=1.74 \pm 0.14 \msun, R=9.3\pm 1.0$ km, \citealt{GO10}),
and \mbox{4U\,1820--30} ($M=1.58 \pm 0.06 \msun, R=9.1\pm 0.4$ km, \citealt{GWCO10}). 
The authors of these papers used only short PRE bursts, which, as we have seen above, are suspicious, 
because their spectral evolution is not consistent with the theory the method is based on, 
with the main reasons being probably partial blocking of the NS by the accretion disk 
and the effects of  the spreading layer on the NS atmosphere.
A high declared accuracy cannot be understood in the light of all the uncertainties, especially on the distance 
and the chemical composition (see Fig.\,\ref{fig:uncrt}). 
We try to find below the answers by the critical consideration of the input numbers and the assumptions made in
the aforementioned papers.

\subsubsection{{EXO\,1745--248} in Terzan~5} 

\begin{figure}
\plotone{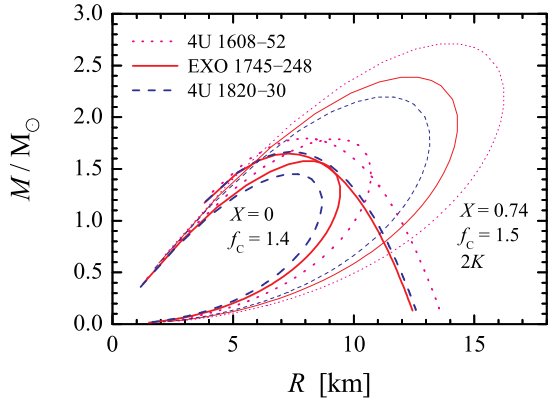}
%\centering 
%\includegraphics[width=\columnwidth]{fig13.eps}
\caption{Curves corresponding to constant $T_{\rm Edd,\infty}$ and $K$ (for the strict lower limit on the distance) 
at $f_{\rm c}$=1.4 and $X$ = 0 for \mbox{EXO\,1745--248} (thick solid curves), \mbox{4U\,1820--30} (thick dashed curves), and 
\mbox{4U\,1608--52} (thick dotted curves). The contours for constant $T_{\rm Edd,\infty}$   
at $f_{\rm c}$=1.5 and $X$ = 0.7374 for double normalization $2K$ are also presented (thin curves). }	
\label{fig:mr_short}
\end{figure}

\citet{OGP09} have determined the following parameters from two PRE bursts from \mbox{EXO\,1745--248}: 
$F_{\rm Edd, -7}$ = 0.625$\pm$0.02 and $K$ = 116$\pm$13. 
These correspond to $T_{\rm Edd,\infty}= 2.2 \times10^7$~K and the maximum possible distance (see Equation (\ref{eq:distmax}))
$D_{10, \max} =0.5978$ at $f_{\rm c}$=1.4 (which was fixed). For the chemical composition, 
the authors also assumed pure helium, $X=0$. 
The distance was taken $D_{10}$ = 0.63$\pm$0.0315 (box-car distribution) 
with the strict lower limit of 0.5985 being very close (within 0.1\%) to  
the maximum possible distance for the observables. 
As a result the curves corresponding to $T_{\rm Edd,\infty}$ and $K$ only touch each other (see Fig.\,\ref{fig:mr_short}). 
We note that the assumption of the box-car distribution for the distance (as well as for $K$) 
only allows distances within 10\% above the used strict lower limit. 
Thus \citet{OGP09} de facto assumed  nearly the delta-function distribution of the distance at $\sim$6 kpc. 
Fixing also $f_{\rm c}$ and $X$ they thus completely removed all uncertainties connected with these parameters. 
The declared errors in $M$ and $R$ thus reflect the statistical errors in $F_{\rm Edd, -7}$ and $K$ 
only, which are of course small because of brightness of the considered events. 
We also note that \citet{Ort07} evaluated the distance to Terzan~5 of 5.5$\pm$0.9 kpc and 
\citet{Valenti07} gives $D=5.9$ kpc. 
Relaxing the assumption of the box-car distribution for the distance (e.g. by using a Gaussian distribution)  
will inevitably move the solutions towards smaller distances and therefore smaller masses and radii. 

\subsubsection{4U\,1820--30 in NGC\,6624}

Using the \rxte\ data for five short PRE bursts from \mbox{4U\,1820--30}, \citet{GWCO10}  determined 
$F_{\rm Edd, -7}$ = 0.539$\pm$0.012 and $K= 92\pm2$.
These correspond to $T_{\rm Edd,\infty} = 2.25\times 10^7$~K 
and the maximum possible distance $D_{10,\max} = 0.617$ at $f_{\rm c}=1.4$ and $X=0$.  
These authors fixed the chemical composition at $X=0$ and varied $f_{\rm c}$ between 1.3 and 1.4 with the strict limits (box-car distribution).
Using estimates by \citet{Kuulkers03} and \citet{Valenti07}, they took 
the distance to the source $D_{10}$ = 0.82$\pm$0.14, but again assumed strict limits (box-car distribution). 

Formally, there is no solution possible for these observables as the curves
corresponding to $T_{\rm Edd,\infty}$ and $K$ do not cross (see Fig.\,\ref{fig:mr_short}).
The solutions for the larger distances in the Gaussian tails of the distributions of $F_{\rm Edd, -7}$ and $K$  are still possible 
with the probability of about $10^{-7}$ \citep{SLB10}. 
Again the obtained errors in $M$ and $R$ reflect the statistical errors in $F_{\rm Edd, -7}$ and $K$ only, but the solution 
is   highly unlikely. 

As in the case of \mbox{EXO\,1745--248}, relaxing the distance constraints to allow smaller distances  
moves the solution towards smaller distances and smaller $M$ and $R$. 
Similarly to the case of the short bursts from \source, the spectral evolution of the considered bursts is not consistent with theory. 
The value of $K^{-1/4}$ drops by only 12\% after touchdown, while the theory predicts 20\% variations.

\subsubsection{4U\,1608--52}

Using the data from four PRE bursts and one non-PRE burst from \mbox{4U\,1608--52}, \citet{GO10} found 
the following observables: $F_{\rm Edd, -7}$ = 1.541$\pm$0.065 and $K$ = 324.6$\pm$2.4. 
These correspond to $T_{\rm Edd,\infty} = 2.13 \times 10^7$~K and the maximum possible distance $D_{10,\max} =0.405$
at $f_{\rm c}$=1.4 and $X$=0. The chemical composition was allowed to vary from $X = 0$ to 0.7, and $f_{\rm c}$ within the interval between  
1.3 and 1.4 with the strict limits. 
The distance was taken by these authors as $D_{10}$ = 0.58$^{+0.2}_{-0.18}$ with 
the strict lower limit equal to 3.9 kpc.  This lower limit in this case is smaller 
than the maximum possible distance and the solutions exist (the curves corresponding to $T_{\rm Edd,\infty}$ and $K$ do cross, 
see Fig.\,\ref{fig:mr_short}). Formally, the authors allow variation of $X$ in a wide range, 
but in reality only solutions with $X \le$ 0.04 are possible. 
At larger $X$ the maximum possible distance becomes smaller than the strict lower limit on the distance of 3.9 kpc.
As in the previous two cases,   
relaxing the distance constraints to allow for smaller distances  
will move the solution towards smaller distances and smaller $M$ and $R$. 

We also note here that bursts selected by \citet{GO10} occurred at a high accretion rate in the soft state. 
The  normalization $K$ is nearly constant during the cooling tails, which strongly contradicts the theory. 
On the other hand, the PRE bursts happening at a low accretion rate follow spectral evolution predicted by the
theory, have normalization $K$ about 60\% larger, 
and, of course, the NS mass and radius determined from these data are also different (Poutanen et al., in preparation).

\subsubsection{Summary} 

We conclude that small uncertainties in $M$ and $R$ obtained for the three bursters 
are the direct consequence of  fixing the color correction $f_{\rm c}$, hydrogen mass fraction $X$, and most importantly 
of the assumed strict lower limit  in the distance distribution, which allows only solutions 
at its lower edge. The lower limits on the distance assumed by the authors
in two cases turned out to be very close to the maximum possible distance allowed by the observables. 
This completely removed all the uncertainties in the distance. Assuming a Gaussian distribution in distance 
allows solutions with much smaller (probably unphysical) $M$ and $R$ for all three sources. 

The results of the spectral fitting for the short PRE bursts for these three NSs 
are very close to the results obtained for the short bursts of \source.
We suggest that during short bursts roughly half of the NS is visible most probably due to the eclipse by the accretion disk. 
Taking the apparent area twice the observed one, 
the determined NS radii move to the values consistent with those determined from the long 
burst from \source\ (see Fig. \ref{fig:mr_short}). 
Additional corrections are also possibly needed because of the 
influence of the boundary layer on the dynamics and the spectral properties of the bursting atmosphere, resulting 
in a higher color correction $f_{\rm c}\sim 1.6$--1.8. 

Recently \citet{SLB10} suggested that the absence of the solutions for short bursts 
mentioned above can be fixed by relaxing the assumption that the Eddington flux is reached at the moment of touchdown
(or rather that the photosphere at touchdown is not at the NS surface).
This, however,  does not solve the problem because the determined blackbody normalization in the cooling tail 
and the Eddington flux are not reliable in these short PRE bursts and the assumed distance constraints were too strict.  
As the spectral evolution during the cooling tails of the short PRE bursts (at high persistent luminosity) 
strongly contradicts the theory, the results obtained from these data are questionable. 
For the determination of the NS masses and radii we suggest  to use only those data that do follow the theory.

\section{Conclusions} 
 
We applied a recently computed detailed set of NS atmosphere models covering a
large range of luminosities \citep{SPW11} 
to the data of the PRE bursts 
of \source.  We showed that the variation of the apparent blackbody radius
during the cooling stage of the 150 s long 
PRE X-ray bursts in \source\ is entirely consistent with the theoretical  color-correction--flux  dependence at intermediate fluxes. 
We thus obtained the Eddington flux and the apparent NS radius (divided by the distance to the source).  
We find that the Eddington flux  (for Thomson opacity) is reached not at the so called ``touchdown'', 
but later at a 15\% lower luminosity. 
We  constrained  mass and radius of the NS using the estimated distance to the source. 
We find a lower limit on the stellar radius of $\sim$14 km for $M<2.3\msun$ at 90\% confidence independently of chemical composition.
(If the NS in \source\ has a spin of 500 Hz, the radius of the non-rotating star would be about 1 km smaller.)
Smaller radii are possible only for more massive NS. 
These results support a stiff equation of state of the NS matter. 
We showed that  hydrogen rich accreting matter is preferred  and  obtained an
upper limit on the distance to \source\ of about 7 kpc. 

We have also demonstrated that the cooling tracks of the short PRE bursts from \source\ 
that occurred during the high/soft state do not follow the evolution expected from theory and the NS apparent areas
are a factor of two smaller than that for the long burst. 
The probable reason is the partial eclipse  of the NS surface by the optically thick accretion disk.
An additional spectral hardening during the cooling tails 
associated with the influence of the boundary/spreading layer can be also important.  
Therefore, the constraints on the NS mass and radius obtained from such bursts are not  reliable. 

Finally, we showed that previous analyses of the short PRE bursts from three sources EXO\,1745--248, 
4U\,1820--30  and 4U\,1608--52 are  questionable, because they ignore the fact that spectral 
evolution during the bursts is not consistent with the  theory for the
passively cooling unobscured NS that is the base for the analysis. Assuming that 
the touchdown flux is reached when the photosphere is detached from the NS surface 
\citep{SLB10} does not solve the problem because the determined blackbody normalization in the cooling tail 
and the Eddington flux are not reliable in these short bursts.   

We suggest that only PRE bursts showing spectral evolution consistent with the theory should be used when 
estimating NS masses and radii.  Further improvement in accuracy of determination of the NS parameters requires 
new atmosphere models with the exact Compton scattering kernel,
as well as accounting for the Doppler effect due to the rapid NS rotation. 
Rotation also introduces additional distortion to the NS spectra, 
because of the difference in the effective gravity at the stellar poles and the equator due to the centrifugal force. 
We plan to investigate the importance of these effects in future work.

\begin{acknowledgements}
The work is supported by the DFG grant SFB / Transregio 7 ``Gravitational Wave Astronomy'' (V.S., K.W.), 
Russian Foundation for Basic Research (grant  09-02-97013-r-povolzhe-a, V.S.),   
the Academy of Finland (grant 127512, J.P.), and DFG cluster of excellence ``Origin and Structure of the Universe" (M.R.).
We thank Dmitry Yakovlev for a number of useful suggestions. 
\end{acknowledgements}

%\bibliographystyle{apj}
%\bibliography{allbib}

\begin{thebibliography}{45}
\expandafter\ifx\csname natexlab\endcsname\relax\def\natexlab#1{#1}\fi

\bibitem[{{Arnaud}(1996)}]{Arn96}
{Arnaud}, K.~A. 1996, in ASP  Conf. Ser. 101, Astronomical Data Analysis Software and Systems V, ed. G.~H.
  {Jacoby} \& J.~{Barnes} (San Francisco: ASP), 17

\bibitem[{{Damen} {et~al.}(1990){Damen}, {Magnier}, {Lewin}, {Tan}, {Penninx},  \& {van Paradijs}}]{Damen90}
{Damen}, E., {Magnier}, E., {Lewin}, W.~H.~G., {Tan}, J., {Penninx}, W., \&
  {van Paradijs}, J. 1990, \aap, 237, 103

\bibitem[{{Ebisuzaki}(1987)}]{Ebi87}
{Ebisuzaki}, T. 1987, \pasj, 39, 287

\bibitem[{{Galloway} {et~al.}(2008{\natexlab{a}}){Galloway}, {Muno}, {Hartman},  {Psaltis}, \& {Chakrabarty}}]{GMH08}
{Galloway}, D.~K., {Muno}, M.~P., {Hartman}, J.~M., {Psaltis}, D., \&
  {Chakrabarty}, D. 2008{\natexlab{a}}, \apjs, 179, 360

\bibitem[{{Galloway} {et~al.}(2008{\natexlab{b}}){Galloway}, {{\"O}zel}, \&  {Psaltis}}]{GOP08}
{Galloway}, D.~K., {{\"O}zel}, F., \& {Psaltis}, D. 2008{\natexlab{b}}, \mnras,
  387, 268

\bibitem[{{Grebel} \& {Roberts}(1995)}]{GR95}
{Grebel}, E.~K., \& {Roberts}, W.~J. 1995, \aaps, 109, 293

\bibitem[{{Guillot} {et~al.}(2011){Guillot}, {Rutledge}, \& {Brown}}]{GRB11}
{Guillot}, S., {Rutledge}, R.~E., \& {Brown}, E.~F. 2011, \apj, 732, 88

\bibitem[{{G{\"u}ver} {et~al.}(2010{\natexlab{a}}){G{\"u}ver}, {{\"O}zel},  {Cabrera-Lavers}, \& {Wroblewski}}]{GO10}
{G{\"u}ver}, T., {{\"O}zel}, F., {Cabrera-Lavers}, A., \& {Wroblewski}, P.
  2010{\natexlab{a}}, \apj, 712, 964

\bibitem[{{G{\"u}ver} {et~al.}(2010{\natexlab{b}}){G{\"u}ver}, {Wroblewski},  {Camarota}, \& {{\"O}zel}}]{GWCO10}
{G{\"u}ver}, T., {Wroblewski}, P., {Camarota}, L., \& {{\"O}zel}, F.
  2010{\natexlab{b}}, \apj, 719, 1807

\bibitem[{{Haensel} {et~al.}(2007){Haensel}, {Potekhin}, \& {Yakovlev}}]{HPY07}
{Haensel}, P., {Potekhin}, A.~Y., \& {Yakovlev}, D.~G. 2007, Astrophysics and
  Space Science Library, Vol. 326, {Neutron Stars 1: Equation of State and
  Structure} (New York: Springer)

\bibitem[{{Hebeler} {et~al.}(2010){Hebeler}, {Lattimer}, {Pethick}, \&  {Schwenk}}]{HLP10}
{Hebeler}, K., {Lattimer}, J.~M., {Pethick}, C.~J., \& {Schwenk}, A. 2010,
  Phys. Rev. Lett., 105, 161102

\bibitem[{{Heinke} {et~al.}(2006){Heinke}, {Rybicki}, {Narayan}, \&  {Grindlay}}]{HR06}
{Heinke}, C.~O., {Rybicki}, G.~B., {Narayan}, R., \& {Grindlay}, J.~E. 2006,
  \apj, 644, 1090
	
\bibitem[{{Inogamov} \& {Sunyaev}(1999)}]{IS99}
{Inogamov}, N.~A., \& {Sunyaev}, R.~A. 1999, Astron. Lett., 25, 269

\bibitem[{{Jahoda} {et~al.}(2006){Jahoda}, {Markwardt}, {Radeva}, {Rots},
  {Stark}, {Swank}, {Strohmayer}, \& {Zhang}}]{JMR06}
{Jahoda}, K., {Markwardt}, C.~B., {Radeva}, Y., {Rots}, A.~H., {Stark}, M.~J.,
  {Swank}, J.~H., {Strohmayer}, T.~E., \& {Zhang}, W. 2006, \apjs, 163, 401

\bibitem[{{Kirsch} {et~al.}(2005){Kirsch}, {Briel}, {Burrows}, {Campana},
  {Cusumano}, {Ebisawa}, {Freyberg}, {Guainazzi}, {Haberl}, {Jahoda},
  {Kaastra}, {Kretschmar}, {Larsson}, {Lubinski}, {Mori}, {Plucinsky},
  {Pollock}, {Rothschild}, {Sembay}, {Wilms}, \& {Yamamoto}}]{Kirsch05}
{Kirsch}, M.~G.,  et al. 2005, Proc. SPIE, 5898, 22 
  
\bibitem[{{Kuchinski} {et~al.}(1995){Kuchinski}, {Frogel}, {Terndrup}, \&  {Persson}}]{1995AJ....109.1131K}
{Kuchinski}, L.~E., {Frogel}, J.~A., {Terndrup}, D.~M., \& {Persson}, S.~E.
  1995, \aj, 109, 1131

\bibitem[{{Kuulkers} {et~al.}(2003){Kuulkers}, {den Hartog}, {in't Zand},  {Verbunt}, {Harris}, \& {Cocchi}}]{Kuulkers03}
{Kuulkers}, E., {den Hartog}, P.~R., {in't Zand}, J.~J.~M., {Verbunt},
  F.~W.~M., {Harris}, W.~E., \& {Cocchi}, M. 2003, \aap, 399, 663

\bibitem[{{Lattimer} \& {Prakash}(2007)}]{LP07}
{Lattimer}, J.~M., \& {Prakash}, M. 2007, \physrep, 442, 109

\bibitem[{{Lewin} {et~al.}(1993){Lewin}, {van Paradijs}, \& {Taam}}]{LvPT93}
{Lewin}, W.~H.~G., {van Paradijs}, J., \& {Taam}, R.~E. 1993, Space Sci.  Rev., 62, 223

\bibitem[{{London} {et~al.}(1984){London}, {Howard}, \& {Taam}}]{London84}
{London}, R.~A., {Howard}, W.~M., \& {Taam}, R.~E. 1984, \apjl, 287, L27

\bibitem[{{London} {et~al.}(1986){London}, {Taam}, \& {Howard}}]{London86}
{London}, R.~A., {Taam}, R.~E., \& {Howard}, W.~M. 1986, \apj, 306, 170

\bibitem[{{Madej}(1991)}]{Madej91}
{Madej}, J. 1991, \apj, 376, 161

\bibitem[{{Madej} {et~al.}(2004){Madej}, {Joss}, \&  {R{\'o}{\.z}a{\'n}ska}}]{MJR04}
{Madej}, J., {Joss}, P.~C., \& {R{\'o}{\.z}a{\'n}ska}, A. 2004, \apj, 602, 904

\bibitem[{{Majczyna} {et~al.}(2005){Majczyna}, {Madej}, {Joss}, \&  {R{\'o}{\.z}a{\'n}ska}}]{MM05}
{Majczyna}, A., {Madej}, J., {Joss}, P.~C., \& {R{\'o}{\.z}a{\'n}ska}, A. 2005,  \aap, 430, 643

\bibitem[{{Molkov} {et~al.}(2000){Molkov}, {Grebenev}, \& {Lutovinov}}]{MGL00}
{Molkov}, S.~V., {Grebenev}, S.~A., \& {Lutovinov}, A.~A. 2000, \aap, 357, L41

\bibitem[{{Ortolani} {et~al.}(2007){Ortolani}, {Barbuy}, {Bica}, {Zoccali}, \&  {Renzini}}]{Ort07}
{Ortolani}, S., {Barbuy}, B., {Bica}, E., {Zoccali}, M., \& {Renzini}, A. 2007,  \aap, 470, 1043

\bibitem[{{Ortolani} {et~al.}(1997){Ortolani}, {Bica}, \& {Barbuy}}]{OBB97}
{Ortolani}, S., {Bica}, E., \& {Barbuy}, B. 1997, \aap, 326, 614

\bibitem[{{{\"O}zel} {et~al.}(2009){{\"O}zel}, {G{\"u}ver}, \&  {Psaltis}}]{OGP09}
{{\"O}zel}, F., {G{\"u}ver}, T., \& {Psaltis}, D. 2009, \apj, 693, 1775

\bibitem[{{Paczynski}(1983)}]{Pac83}
{Paczynski}, B. 1983, \apj, 267, 315

\bibitem[{{Pavlov} {et~al.}(1991){Pavlov}, {Shibanov}, \& {Zavlin}}]{PSZ91}
{Pavlov}, G.~G., {Shibanov}, I.~A., \& {Zavlin}, V.~E. 1991, \mnras, 253, 193

\bibitem[{{Penninx} {et~al.}(1989){Penninx}, {Damen}, {van Paradijs}, {Tan}, \&  {Lewin}}]{Penninx89}
{Penninx}, W., {Damen}, E., {van Paradijs}, J., {Tan}, J., \& {Lewin}, W.~H.~G. 1989, \aap, 208, 146

\bibitem[{{Pozdniakov} {et~al.}(1983){Pozdniakov}, {Sobol}, \&  {Sunyaev}}]{PSS83}
{Pozdniakov}, L.~A., {Sobol}, I.~M., \& {Sunyaev}, R.~A. 1983, Astrophys. and  Space Phys. Rev., 2, 189
  
\bibitem[{{Rutledge} {et~al.}(2002){Rutledge}, {Bildsten}, {Brown}, {Pavlov},  \& {Zavlin}}]{R02}
{Rutledge}, R.~E., {Bildsten}, L., {Brown}, E.~F., {Pavlov}, G.~G., \& {Zavlin}, V.~E. 2002, \apj, 577, 346

\bibitem[{{Steiner} {et~al.}(2010){Steiner}, {Lattimer}, \& {Brown}}]{SLB10}
{Steiner}, A.~W., {Lattimer}, J.~M., \& {Brown}, E.~F. 2010, \apj, 722, 33

\bibitem[{{Strohmayer} \& {Bildsten}(2006)}]{SB06}
{Strohmayer}, T., \& {Bildsten}, L. 2006, in   Cambridge Astrophysics Series 39, 
Compact stellar X-ray sources, ed. W.~{Lewin} \& M.~{van der Klis}
  (Cambridge: Cambridge University Press), 113

\bibitem[{{Suleimanov} {et~al.}(2006){Suleimanov}, {Madej}, {Drake}, {Rauch},  \& {Werner}}]{SMD06}
{Suleimanov}, V., {Madej}, J., {Drake}, J.~J., {Rauch}, T., \& {Werner}, K. 2006, \aap, 455, 679

\bibitem[{{Suleimanov} \& {Poutanen}(2006)}]{SulP06}
{Suleimanov}, V., \& {Poutanen}, J. 2006, \mnras, 369, 2036

\bibitem[{{Suleimanov} {et~al.}(2011){Suleimanov}, {Poutanen}, \&  {Werner}}]{SPW11}
{Suleimanov}, V., {Poutanen}, J., \& {Werner}, K. 2011, \aap, 527, A139 

\bibitem[{{Suleimanov} \& {Werner}(2007)}]{SW07}
{Suleimanov}, V., \& {Werner}, K. 2007, \aap, 466, 661

\bibitem[{{Valenti} {et~al.}(2007){Valenti}, {Ferraro}, \&  {Origlia}}]{Valenti07}
{Valenti}, E., {Ferraro}, F.~R., \& {Origlia}, L. 2007, \aj, 133, 1287

\bibitem[{{van Paradijs} {et~al.}(1990){van Paradijs}, {Dotani}, {Tanaka}, \&  {Tsuru}}]{vP90}
{van Paradijs}, J., {Dotani}, T., {Tanaka}, Y., \& {Tsuru}, T. 1990, \pasj, 42, 633

\bibitem[{{Webb} \& {Barret}(2007)}]{WB07}
{Webb}, N.~A., \& {Barret}, D. 2007, \apj, 671, 727

\bibitem[{{Weisskopf} {et~al.}(2010){Weisskopf}, {Guainazzi}, {Jahoda},
  {Shaposhnikov}, {O'Dell}, {Zavlin}, {Wilson-Hodge}, \&  {Elsner}}]{Weisskopf10}
{Weisskopf}, M.~C., {Guainazzi}, M., {Jahoda}, K., {Shaposhnikov}, N.,
  {O'Dell}, S.~L., {Zavlin}, V.~E., {Wilson-Hodge}, C., \& {Elsner}, R.~F.
  2010, \apj, 713, 912
  
\bibitem[{{Zane} {et~al.}(2000){Zane}, {Turolla}, \& {Treves}}]{ZTT00}
{Zane}, S., {Turolla}, R., \& {Treves}, A. 2000, \apj, 537, 387
  
\bibitem[{{Zavlin} {et~al.}(1996){Zavlin}, {Pavlov}, \& {Shibanov}}]{ZPS96}
{Zavlin}, V.~E., {Pavlov}, G.~G., \& {Shibanov}, Y.~A. 1996, \aap, 315, 141

\bibitem[{{Zhang} {et~al.}(2010){Zhang}, {Mendez}, \& {Altamirano}}]{ZMA10}
{Zhang}, G., {Mendez}, M., \& {Altamirano}, D. 2010, \mnras, 413, 1913 

\end{thebibliography}

%\clearpage

%\clearpage

\end{document}